\documentclass{article}
\usepackage{smc2017}
\usepackage{times}
\usepackage{ifpdf}
\usepackage[english]{babel}
\usepackage{cite}
\usepackage{multirow}
\usepackage{subcaption}
\usepackage{url}
\usepackage{color,soul}
\usepackage{flushend}

\def\papertitle{Stacked Convolutional and Recurrent Neural Networks for Music Emotion Recognition}
\def\firstauthor{Miroslav Malik}
\def\secondauthor{Sharath Adavanne, Konstantinos Drossos, Tuomas Virtanen, Dasa Ticha, Roman Jarina}

\newif\ifpdf
\ifx\pdfoutput\relax
\else
   \ifcase\pdfoutput
      \pdffalse
   \else
      \pdftrue
\fi

\ifpdf 
  \usepackage[pdftex,
    pdftitle={\papertitle},
    pdfauthor={\firstauthor, \secondauthor},
    bookmarksnumbered, 
    pdfstartview=XYZ 
   ]{hyperref}

  \usepackage[pdftex]{graphicx}
  \graphicspath{{./figures/}}
  \DeclareGraphicsExtensions{.pdf,.jpeg,.png}

  \usepackage[figure,table]{hypcap}

\else 
  \usepackage[dvips,
    bookmarksnumbered, 
    pdfstartview=XYZ 
  ]{hyperref}  

  \usepackage[dvips]{epsfig,graphicx}
  \graphicspath{{./figures/}}
  \DeclareGraphicsExtensions{.eps}

  \usepackage[figure,table]{hypcap}
\fi

\hypersetup{
    colorlinks,%
    citecolor=black,%
    filecolor=black,%
    linkcolor=black,%
    urlcolor=black
}

\title{Stacked Convolutional and Recurrent Neural Networks\\for Music Emotion Recognition}

\author{
 \bf{Miroslav Malik\textsuperscript{*\textdaggerdbl}}, \bf{Sharath Adavanne\textsuperscript{*\textsection}}, \bf{Konstantinos Drossos\textsuperscript{\textsection}}, \bf{Tuomas Virtanen\textsuperscript{\textsection}}, \bf{Dasa Ticha\textsuperscript{\textdaggerdbl}}, and \bf{Roman Jarina\textsuperscript{\textdaggerdbl}}\\
\textsuperscript{\textdaggerdbl}Department of Multimedia and Information-Communication Technologies, University of Zilina, Slovakia\\  \texttt{firstname.lastname@fel.uniza.sk}\\
\textsuperscript{\textsection}Audio Research Group, Department of Signal Processing, Tampere University of Technology, Finland\\
\texttt{firstname.lastname@tut.fi}
\thanks{\textsuperscript{*}Equally contributing authors in this paper.}\thanks{The research leading to these results has received funding from the European Research Council under the European Union’s H2020 Framework Programme through ERC Grant Agreement 637422 EVERYSOUND. Part of the computations leading to these results were performed on a TITAN-X GPU donated by NVIDIA. The authors also wish to acknowledge CSC-IT Center for Science, Finland, for computational resources.} 
}

\begin{document}
\capstartfalse
\maketitle
\capstarttrue
\begin{abstract}
This paper studies the emotion recognition from musical tracks in the 2-dimensional valence-arousal (V-A) emotional space. We propose a method based on convolutional (CNN) and recurrent neural networks (RNN), having significantly fewer parameters compared with state-of-the-art method for the same task. We utilize one CNN layer followed by two branches of RNNs trained separately for arousal and valence. The method was evaluated using the ``MediaEval2015 emotion in music'' dataset. We achieved an RMSE of 0.202 for arousal and 0.268 for valence, which is the best result reported on this dataset.
\end{abstract}

\section{Introduction}\label{sec:introduction}
Music has been used to transfer and convey emotions since the initial era of human communication~\cite{juslin:2005:chapter, juslin:2003:pshycb}. In the field of music signal processing, either perceived or induced emotion (see~\cite{juslin2004expression} for details) has been proven to be a strong and content-oriented attribute of music and has been employed for categorization tasks. For example, published works focus on emotion recognition from music in order to provide an alternative and content-oriented categorization scheme to the typical ``Artist{\slash}Band{\slash}Year'' one~\cite{zhang:2008:icme, li:2004:icassp}.  

There are two main categories of models for emotions; i) the discrete, and ii) the dimensional models. Discrete models of emotion employ a discrete group of words to model emotions. The basic emotions model~\cite{ekman:1992:anargument} and the Hevner's list~\cite{henver:1936:henverlist} are the important representative examples of this group. The main drawback of discrete models of emotion is the subjective perception of the meaning of each word that is employed to describe an emotion. This leads to confusion when different words are utilized for the same emotion, e.g. ``Joy''--``Enjoyment'' and ``Cheerfulness''--``Happiness''~\cite{juslin:2003:pshycb, drossos:2013:iisa}. Despite this confusion, discrete models are heavily used in works where the focus is on a specific emotion or emotional condition~\cite{drossos:2012:am}. 

This categorical approach is also employed in the audio mood classification (AMC) task, under the music information retrieval evaluation exchange (MIREX)~\cite{downie20082007}. At AMC, the participants were requested to propose methods and systems capable of classifying audio files into five discrete emotion classes, using one class for the whole audio clip. The five classes contained five to seven adjectives and these adjectives were derived from online tags and were not based on some of the accepted categorical models of emotion. On average, the accuracy of the proposed methods and systems was below 70\%~\cite{mirex2016,mirex2011}.

Dimensional models represent emotion in a continuous $N$-dimensional space. Usually, $N=2$ and one dimension corresponds to arousal and the other to valence~\cite{drossos:2015:taffc}. The emotion in a song is usually not spread equally through the time and it varies around some bias value. The bias value usually stays unchanged for a significant fraction of the duration of the song~\cite{xianyu2016svr}. This property can be hardly observable in the categorical approach, but with the dimensional approach, the continuous changes can be captured more easily.

\sloppy
A dimensional model was employed for the emotion in music (EiM) challenge, organized under the MediaEval benchmark~\cite{aljanaki2014emotion}. To the best of authors' knowledge, MediaEval along with MIREX are the two widely known international scientific challenges for music emotion recognition. The MediaEval challenge consisted of two tasks, one focusing on static and the other focused on the dynamic presence of emotion in music. In the second (dynamic) task of the EiM, the methods proposed were based on support vector regression (SVR)~\cite{chmulik2015uniza}, continuous conditional neural fields~\cite{imbrasaite2014music}, continuous conditional random fields~\cite{cai2015pku}, feed-forward neural networks~\cite{patra2015mediaeval}, and recurrent neural networks~\cite{pellegrini2015time}. The dataset for this dynamic task of EiM consisted of extracted audio features and emotional annotations. The features and the annotations were for frames of audio of length 500 ms. The best-reported results for this task were obtained using an ensemble of six long short-term memory (LSTM) RNNs with different input sequence lengths. The final output was predicted from the ensemble using an extreme learning machine. This method achieved a root-mean-square error (RMSE) of 0.230 for arousal and 0.303 for valence~\cite{xu2015multi}.

In this paper, we propose a method that exploits the combined modeling capacities of the CNN, RNN, and fully connected (FC) network. We call this the stacked convolutional and recurrent neural network (CRNN). Similar architecture of stacked CNN, RNN, and FC has been used in literature and shown to consistently outperform the previous network configurations for sound event detection~\cite{Adavanne2017}, speech recognition~\cite{sainath2015}, and music classification~\cite{Choi2016}. We approach the problem as a regression one and evaluate the method using the EiM dynamic task dataset. We use RMSE as the metric in order to provide a direct comparison with results from existing works. Our proposed method employs only a small fraction of parameters compared to the recent state-of-the-art DBLSTM-based system by Li at al.~\cite{li2016dblstm} and performs with significantly better results (i.e. lower RMSE for both arousal and valence recognition). Additionally, we evaluate our method with a raw feature set (log mel-band energy). This is motivated from the fact that neural networks can by themselves learn the first order derivatives and statistics existing in the feature set provided in EiM dataset. These raw features are extracted from the same songs that were utilized in the EiM dataset. The results using the raw feature set are comparable to the ones achieved with the Li et al. system. 

The rest of the paper is organized as follows. In Section~\ref{sec:crnn} we present our proposed method. Section~\ref{sec:eval} describes the dataset, features, metric, baseline and the evaluation procedure. We analyze and discuss the obtained results in Section~\ref{sec:results}. Finally concluding the paper in Section~\ref{sec:conclusions}.

\section{Proposed method}\label{sec:crnn}

\begin{figure}
\centering
\includegraphics[width=\columnwidth]{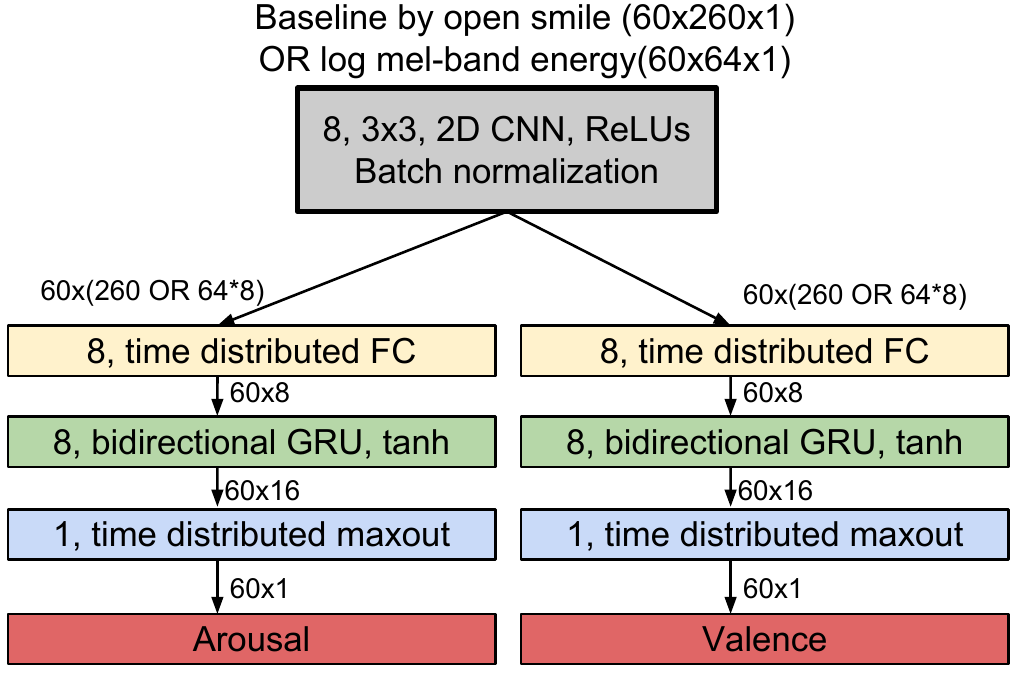}
\caption{Proposed method of stacked convolutional and recurrent neural network for music emotion recognition.}
\label{fig:crnn}
\end{figure}

The input to the proposed method is a sequence of audio features extracted from audio. We evaluate the method with two separate audio features - baseline (Section \ref{ssec:baselineFeat}) and raw (Section \ref{ssec:secondaryFeat}). Our proposed method, illustrated in Figure \ref{fig:crnn} is a neural network architecture obtained by stacking a CNN with two branches of FC and RNN one each for arousal and valence. This combined CRNN architecture maps the input audio features into their respective arousal and valence values. The output of the method, arousal and valence values, is in the range of [-1, 1].

In our method, the local shift-invariant features are extracted from the audio features using a CNN with a receptive filter size of $3\times3$. The feature map of the CNN acts as the input to two parallel but identical branches, one for arousal and the other one for valence. Each of these branches consists of the FC, the bidirectional gated recurrent unit (GRU)~\cite{cho2014learning}, and the output layer consisting of one node of the maxout layer~\cite{Goodfellow2013}. Both FC and maxout had their weights shared across time steps. The FC was utilized to reduce the dimensionality of the feature maps from the CNN and, consequently, reduce the number of parameters required by the GRU. The bidirectional GRU layer was utilized to learn the temporal information in the features. The maxout layer was employed as the regression layer due to its ability to approximate a convex, piecewise linear activation function~\cite{Goodfellow2013}. 

We use the rectified linear unit (ReLU)~\cite{nair2010rectified} activation and the batch normalization \cite{batchNorm} for the CNN layer. The FC uses the linear activation and the GRU's use the tanh activation. In the bidirectional GRU, we concatenate together the forward and backward activations. The network was trained using the backpropagation through time alorithm \cite{bptt1990}, the Adam optimizer with the default parameters in~\cite{adamKeras}, and the RMSE as the loss function. In order to reduce overfitting of the network to training data we use dropout~\cite{Dropout} for the CNN and RNN layers and the ElasticNet~\cite{Zou2005} regularization for the weights and activity of the CNN. The network was implemented using the Keras framework~\cite{chollet2015keras} and the Theano backend~\cite{bergstra2010theano}. 

\section{Evaluation}
\label{sec:eval}
\subsection{Dataset}
\label{ssec:data}
For the evaluation of our method, we utilized the dataset provided for the MediaEval's EiM task~\cite{aljanaki2014emotion}. For annotations, the Russell's two-dimensional continuous emotional space~\cite{russell1980circumplex} was used. The values of arousal and valence belonging to every 500 ms segment were annotated by five to seven annotators per song from the Amazon Mechanical Turk.

The training set consisted of 431 audio excerpts with a length of 45 seconds from the Free music archive. The first 15 seconds of excerpts were used for annotator's adaptation. The annotations and features for the remaining 30 seconds were provided as the development set. This amounts to 60 annotations per audio file (30 s of audio with annotations every 500 ms). Each of the annotations are in the range of [-1, 1] for both arousal and valence. Negative values represents low arousal/valence and positive high.

The evaluation set consists of 58 full songs from the royalty free multitrack MedleyDB dataset and Jamendo music website. Similar features and annotations as development set were provided for the evaluation set.

\subsection{Audio features}

\subsubsection{Baseline features}
\label{ssec:baselineFeat}
The feature set that we used is the one that was utilized in the EiM challenge. The most of research is based on this feature set, so we call it the baseline features set. It consists of the mean and standard deviation of 65 low-level acoustic descriptors and their first order derivatives from the 2013 INTERSPEECH Computational Paralinguistic Challenge~\cite{weninger2013acoustics}, extracted with the openSMILE toolbox~\cite{eyben2013recent}. In the feature set are included mel frequency cepstral coefficients (MFCCs), spectral features such as flux, centroid, kurtosis, and rolloff, and voice related features such as jitter and shimmer. The total amount of features is 260 and they were extracted from non overlapping segments of length 500 ms. The mean and standard deviation of the features for this 500 ms were estimated from frames of 60 ms with 50 ms ($\approx80\%$) overlap. 

\subsubsection{Raw audio feature}
\label{ssec:secondaryFeat}
The above baseline features consist of mean, standard deviation, first and second order derivatives of different raw features. Neural networks by themselves can extract such statistics from the raw features directly. Thus, we argue that these features are not the best choice for neural networks. In order to study the performance of the proposed network with raw features, we employ just the log mel-band energies. Mel-band related features have been used previously for MediaEval EiM task~\cite{cai2015pku} and MIREX AMC task~\cite{lidy2016parallel}. The librosa python library~\cite{librosa} was used to extract the mel-band features from 500 ms segments, in a similar fashion to the baseline feature set.

\subsection{Metric}
The RMSE is widely used in many areas as a statistical metric to measure a model performance. It represents a standard deviation of the differences of predicted values from the line of best fit at the sample level. Given \(N\) predicted samples \(\hat{y}_n\) and the corresponding reference samples \(y_n\), then RMSE between them can be written as:

\begin{equation} 
RMSE = \sqrt{\frac{\sum_{n=1}^{N} (\hat{y}_n - y_n)^2} {N}} \label{eq:4.1}
\end{equation}

\subsection{Baseline}
The proposed method is compared to the system proposed in \cite{li2016dblstm}. To the best of our knowledge, this method has the top results on the MediaEval's EiM dataset using the baseline audio features. This result was obtained using an ensemble of six bidirectional LSTM (BLSTM) networks of five layers and 256 units each, trained on sequence length of 10. The first two layers of these were pre-trained with the baseline features. The six networks of the ensemble were chosen such that they covered all the training data, and sequence lengths of 10, 20, 30 and 60. The output of the ensemble of networks was fused using SVR with a third order polynomial kernel and artificial neural network (ANN) of 14 nodes. The average output of SVR and ANN was used as the final arousal and valence values.

In terms of complexity of this baseline method, a five layer 256 input and output units BLSTM has about 2 million parameters. An ensemble of six BLSTMs compounded in this manner will have about 12 million parameters. Additionally, SVR and ANN adds to further complexity of the overall method.

\begin{figure}
\centering
\includegraphics[width=0.8\columnwidth]{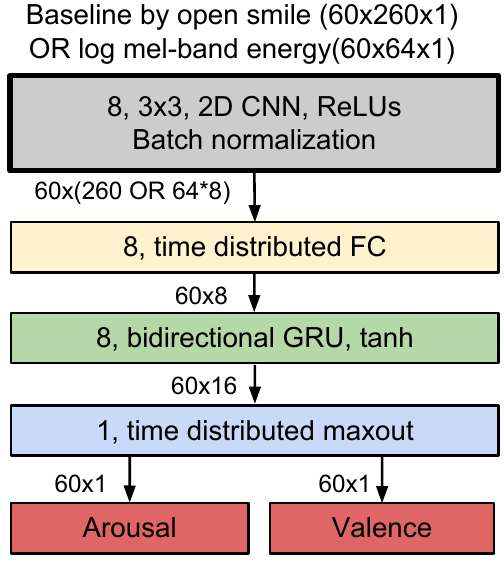}
\caption{Stacked convolutional and recurrent neural network without branching (CRNN\_NB).}
\label{fig:crnn_nobranch} \vspace{-5pt}
\end{figure}

\begin{table*}[!ht]
\caption{RMSE errors for development and evaluation data using (a) baseline and (b) log mel-band energy features. The numbers reported are the mean and standard deviation over five separate runs of the method with different initialization of network weights.. The proposed CRNN network performance is reported for sequence lengths 10, 20, 30 and 60.}
\label{Table:1}
\begin{subtable}{0.95\textwidth}
\centering
\caption{Baseline features}
\label{table:baseline}
\resizebox{0.95\textwidth}{!}{
\begin{tabular}{c|ccc|ccc}
                                                                                       & \multicolumn{3}{c|}{Evaluation} & \multicolumn{3}{c}{Development} \\\cline{2-7}
                                                                                    Seq. length  & Valence   & Arousal  & Average  & Valence  & Arousal  & Average  \\\hline

60 & 0.275$\pm$0.004     & 0.210$\pm$0.004    & 0.242$\pm$0.002    & 0.065$\pm$0.002    & 0.055$\pm$0.001    & 0.060$\pm$0.001    \\

30 & 0.268$\pm$0.004     & 0.205$\pm$0.004    & 0.236$\pm$0.001    & 0.060$\pm$0.001    & 0.053$\pm$0.002    & 0.057$\pm$0.001    \\

20 & 0.268$\pm$0.003     & \bf 0.202$\pm$0.007    & \bf 0.235$\pm$0.002    & \bf 0.059$\pm$0.001    & \bf 0.050$\pm$0.001    & \bf 0.055$\pm$0.001    \\

10 & \bf 0.267$\pm$0.003     & 0.203$\pm$0.006    & \bf 0.235$\pm$0.003    & \bf 0.059$\pm$0.002   & 0.052$\pm$0.001    & \bf 0.055$\pm$0.001   
\end{tabular}
}
\end{subtable}
\linebreak\linebreak\linebreak
\begin{subtable}{0.95\textwidth}
\centering
\caption{Log mel-band energy feature}
\label{table:mel}
\resizebox{0.95\textwidth}{!}{
\begin{tabular}{c|ccc|ccc}
   & \multicolumn{3}{c|}{Evaluation} & \multicolumn{3}{c}{Development} \\\cline{2-7}
Seq. length   & Valence   & Arousal  & Average  & Valence  & Arousal  & Average \\\hline
60 & 0.276$\pm$0.006     & 0.252$\pm$0.009    & 0.264$\pm$0.001    & 0.083$\pm$0.002    & 0.073$\pm$0.002    & 0.078$\pm$0.001    \\
30 & \bf 0.270$\pm$0.003     & 0.250$\pm$0.004    & 0.260$\pm$0.001    & 0.079$\pm$0.001    & 0.070$\pm$0.001    & 0.074$\pm$0.001    \\
20 & \bf 0.270$\pm$0.003     & 0.248$\pm$0.004    & 0.259$\pm$0.002    & 0.079$\pm$0.001    & 0.069$\pm$0.001    & 0.074$\pm$0.000    \\
10 & 0.273$\pm$0.008     & \bf 0.244$\pm$0.006    & \bf 0.258$\pm$0.002    & \bf 0.078$\pm$0.000    & \bf 0.067$\pm$0.001    & \bf 0.072$\pm$0.000   
\end{tabular}
}
\end{subtable}
\end{table*}

\subsection{Evaluation procedure}
The hyperparameter estimation of the proposed method was done by varying the number of layers of the CNN, FC, and GRU from one to three, and the number of units for each of these were varied in the set of $\{4, 8, 16, 32\}$. Identical dropout rates were used for CNN and GRU and varied in the set of $\{0.25, 0.5, 0.75\}$. The ElasticNet variables L1 and L2 were each varied in the set of $\{$1, 0.1, 0.01, 0.001, 0.0001$\}$. The parameters were decided based on the best RMSE score on the development set, using the baseline features, mini-batch size of 32, and the maximal sequence length of 60. The mini-batch size of 32 was chosen from the set of $\{16, 32, 64, 128\}$ based on the variance in the training set error and the number of iterations taken to achieve the best RMSE. The sequence length was chosen to be the same length as the number of annotations per audio file in the training set, therefore 30 s per every audio clip annotated every 0.5 s and this sequence length varied from 10 s to 60 s with the multiply of 2. The network was trained to achieve the lowest average RMSE of valence and arousal. The best configuration with least number of parameters had one layer each of CNN, FC and GRU with eight units each, a dropout rate of 0.25, L1 of 0.1 and L2 of 0.001. Figure~\ref{fig:crnn} shows the configuration and the feature map dimensions of the proposed method. 

In order to provide the comparison of the performance of CRNN without these two branches (CRNN\_NB), we removed one branch in the above configuration and trained the CRNN with both the valence and arousal on the same branch as seen in Figure~\ref{fig:crnn_nobranch}.

The proposed CRNN method was trained with the raw features (log mel-band energy). In order to give a direct comparison of performance, we keep the exact configuration as the proposed CRNN with only the dropout changed to 0.75. The network was seen to overfit to the training data with 0.25 dropout rate that used for baseline features. A dropout rate of 0.75 was chosen after tuning in the set of $\{0.25, 0.5, 0.75\}$. 

\vspace{15pt}
\section{Results and Discussion}\label{sec:results}
\vspace{5pt}
Tables~\ref{table:baseline} and~\ref{table:mel} present the RMSE scores on the development and evaluation datasets for the baseline and log mel-band features, respectively. These scores are the mean and standard deviation over five separate runs of the method with different initialization of network weights. 

The proposed CRNN method with baseline features gave an average RMSE of 0.242 on evaluation set (see average RMSE for sequence length 60 in Table~\ref{table:baseline}). The Li et al. baseline method gave an average RMSE of 0.255 (see Table~\ref{Table:2}). In comparison, the CRNN method has only about 30 k parameters (about 400 times fewer) and fares significantly better in RMSE sense than the Li et al. system. Potentially, further improvement can be achieved by using an ensemble of the proposed method covering different sequence lengths and all training data as proposed in the Li et al. system.

The configuration of the CRNN\_NB method had about 17 k parameters and gave the same average RMSE as the Li et al. system (see CRNN\_NB in Table \ref{Table:2}). This shows the robustness of the proposed method for the emotion recognition task.

Different sequence lengths were experimented with the proposed CRNN, and the results are presented in Table \ref{Table:1}. We use a maximum length of 60 which is equal to the number of annotations from a single file in development set, and its factors of 10, 20 and 30. We see that lower sequence lengths of 10 and 20 perform better than using the full-length sequence of 60. Similar observation was reported in \cite{li2016dblstm}. The best average RMSE of 0.235 was achieved with sequences 10 and 20, this is the best result achieved in this paper and is 0.02 less than Li et al.

We compare the performance of the proposed CRNN with log mel-band features in Table~\ref{table:mel}. Due to the reduced number of features, this network has only about 10 k parameters. With a simple and not so finely tuned network and raw features we get a best average RMSE of 0.258 with sequence length 10. This is very close to the Li et al. system performance with 1200 times fewer parameters. This proves our hypothesis that the neural network can learn the information of first and second order derivatives and first order statistics from the raw features on its own. The trend of shorter sequences performing better than longer sequences is also observed with log mel-band energy. A network tuned for log mel-band features specifically has a potential to perform better than the Li et al. system.

\begin{table}
\centering
\caption{RMSE errors for evaluation data using baseline features. The numbers reported for CRNN\_NB are the mean and standard deviation over five separate runs of the method with different initialization of network weights.}
\resizebox{0.5\textwidth}{!}{
\begin{tabular}{l|lll}				
Method                       & Valence   & Arousal  & Average       \\\hline
Li et al.~\cite{li2016dblstm}        &    0.285     & 0.225    & 0.255      \\
CRNN\_NB                & 0.279$\pm$0.004     & 0.231$\pm$0.003    & 0.255$\pm$0.002     \\
\end{tabular}
}
\label{Table:2}
\end{table}

\vspace{-15pt}
\section{Conclusion}\label{sec:conclusions}
In this paper, we proposed a method consisting of stacked convolutional and recurrent neural networks for continuous prediction of emotion in two-dimensional V-A space. The proposed method used significantly less amount of parameters than the Li et al. system and the obtained results outperform those of the Li et al. system. 

The proposed CRNN was evaluated with different lengths of sequences, and the smaller sequence lengths were seen to perform better than the longer lengths. Log mel-band energy feature was proposed in place of baseline features, and the proposed CRNN was seen to learn information equivalent to that of baseline features from just the mel-band features.

\bibliography{smc2017template}
\end{document}